\documentclass[preprint,superscriptaddress]{revtex4}
\usepackage{graphicx}

\newcommand{\ket}[1] {\left| #1 \right\rangle}

\begin{document}

\title{Extremely large Lamb shift in a deep-strongly coupled circuit QED system with a multimode resonator}

\author{Ziqiao Ao}
\affiliation{Department of Applied Physics, Waseda University, Okubo 3-4-1, Shinjuku-ku, Tokyo 169-8555, Japan}
\affiliation{Department of Advanced Science and Engineering, Waseda University, 3-4-1 Okubo, Shinjuku-ku, Tokyo 169-8555, Japan}
\affiliation{Advanced ICT Institute, National Institute of Information and Communications Technology, 4-2-1, Nukuikitamachi, Koganei, Tokyo 184-8795, Japan}
\author{Sahel Ashhab}
\affiliation{Advanced ICT Institute, National Institute of Information and Communications Technology, 4-2-1, Nukuikitamachi, Koganei, Tokyo 184-8795, Japan}
\author{Fumiki Yoshihara}
\affiliation{Advanced ICT Institute, National Institute of Information and Communications Technology, 4-2-1, Nukuikitamachi, Koganei, Tokyo 184-8795, Japan}
\affiliation{Department of Physics, Tokyo University of Science, 1-3 Kagurazaka, Shinjuku-ku, Tokyo 162-8601, Japan}
\author{Tomoko Fuse}
\affiliation{Advanced ICT Institute, National Institute of Information and Communications Technology, 4-2-1, Nukuikitamachi, Koganei, Tokyo 184-8795, Japan}
\author{Kosuke Kakuyanagi}
\affiliation{NTT Basic Research Laboratories, NTT Corporation, 3-1 Morinosato-Wakamiya, Atsugi, Kanagawa 243-0198, Japan}
\author{Shiro Saito}
\affiliation{NTT Basic Research Laboratories, NTT Corporation, 3-1 Morinosato-Wakamiya, Atsugi, Kanagawa 243-0198, Japan}
\author{Takao Aoki}
\affiliation{Department of Applied Physics, Waseda University, Okubo 3-4-1, Shinjuku-ku, Tokyo 169-8555, Japan}
\author{Kouichi Semba}
\affiliation{Advanced ICT Institute, National Institute of Information and Communications Technology, 4-2-1, Nukuikitamachi, Koganei, Tokyo 184-8795, Japan}
\affiliation{Institute for Photon Science and Technology, The University of Tokyo, 7-3-1 Hongo, Bunkyo-ku, Tokyo 113-0033, Japan}


\begin{abstract}

We report experimental and theoretical results on the extremely large Lamb shift in a multimode circuit quantum electrodynamics (QED) system in the deep-strong coupling (DSC) regime, where the qubit-resonator coupling strength is comparable to or larger than the qubit and resonator frequencies.
The system comprises a superconducting flux qubit (FQ) and a quarter-wavelength coplanar waveguide resonator ($\lambda/4$ CPWR) that are coupled inductively through a shared edge that contains a Josephson junction to achieve the DSC regime.
Spectroscopy is performed around the frequency of the fundamental mode of the CPWR, and the spectrum is fitted by the single-mode quantum Rabi Hamiltonian to obtain the system parameters.
Since the qubit is also coupled to a large number of higher modes in the resonator, the single-mode fitting does not provide the bare qubit energy but a value that incorporates the renormalization from all the other modes.
We derive theoretical formulas for the Lamb shift in the multimode resonator system.
As shown in previous studies, there is a cut-off frequency $\omega_{\rm{cutoff}}$ for the coupling between the FQ and the modes in the CPWR, where the coupling grows as $\sqrt{\omega_n}$ for $\omega_n/\omega_{\rm{cutoff}}\ll 1$ and decreases as $1/\sqrt{\omega_n}$ for $\omega_n/\omega_{\rm{cutoff}}\gg 1$.
Here $\omega_n$ is the frequency of the $n$th mode.
The cut-off effect occurs because the qubit acts as an obstacle for the current in the resonator, which suppresses the current of the modes above $\omega_{\rm{cutoff}}$ at the location of the qubit and results in a reduced coupling strength.
Using our observed spectrum and theoretical formulas, we estimate that the Lamb shift from the fundamental mode is 82.3\% and the total Lamb shift from all the modes is 96.5\%.
This result illustrates that the coupling to the large number of modes in a CPWR yields an extremely large Lamb shift but does not suppress the qubit energy to zero, which would happen in the absence of a high-frequency cut-off.
\end{abstract}

\maketitle

\section*{Introduction}

The interaction between an atom and an electromagnetic (EM) field has been actively studied not only to understand novel quantum physics phenomena but also to develop quantum communication and information processing technologies \cite{NielsenChuang10th,KimbleNature2008,DevoretSchoelkopfScience2013}.
One well-known and well-studied phenomenon caused by this interaction is the Lamb shift \cite{LambPhysRev1947}, where the energy of the atom is slightly renormalized by the coupling with the vacuum fluctuations, which have so-called "half-photon" energy \cite{BrunePRL1994, FragnerScience2008}.
When the coupling strength is pushed into stronger regimes \cite{Forn-DiazRevModPhys2019, KockumNRevPhys2019} such as the ultra-strong coupling (USC) or deep-strong coupling (DSC) regimes, the Lamb shift is no longer a small correction but a dominant contribution that drastically changes the atomic energy \cite{YoshiharaNPhys2017,YoshiharaPRA2017}.

Among several systems that can reach the USC and DSC regimes \cite{AnapparaPRB2009,GambinoACSPhoto2014,BenzScience2016,BayerACSPhoto2017,VrajitoareaarXiv2022}, superconducting quantum circuits (SQC) \cite{BlaisPRA2004,WallraffNature2004,ChiorescuNature2004} are well suited for investigating the Lamb shift, because the USC and DSC regimes can be achieved using a single (artificial) atom (or qubit) coupled to either a single-mode or a multimode superconducting resonator \cite{BourassaPRA2009,NiemczykNPhys2010,LizuainPRA2010,Forn-DiazPRL2010,FelicettiPRA2014,YoshiharaNPhys2017,Forn-DiazNatPhys2017,YoshiharaPRA2017}.
In particular, multimode systems are attractive for the study of novel phenomena such as many-body effects \cite{SundaresanPRX2015,LiuNatPhys2017,BosmannpjQI2017,MartineznpjQI2019} and photon frequency conversion \cite{KoshinoPRResearch2022}.
Understanding the Lamb shift in a multimode DSC system is an important step to establishing better control over such a complex system.

In this work, to study the Lamb shift caused by the coupling between an artificial atom and a multimode resonator in the DSC regime, we investigated circuit QED systems where an FQ is coupled to a $\lambda/4$ CPWR.
The qubit and the resonator are coupled inductively through a shared-edge with a Josephson junction inserted to induce a large inductance and enhance the coupling strength in order to reach the DSC regime \cite{NiemczykNPhys2010, YoshiharaNPhys2017, YoshiharaPRL2018, YoshiharaSciRep2022}.
We conducted spectroscopy experiment on our system around the fundamental mode frequency.
By using the single-mode quantum Rabi model (QRM), we can fit the spectrum with the QRM Hamiltonian and obtain the system parameters.
The fitting provides the qubit's bare frequency, the resonator frequency, and the coupling strength. The obtained qubit frequency is a renormalized one rather than the truly bare one, the reason being the renormalization from all the other modes in the resonator that are not included explicitly in our single-mode QRM Hamiltonian.
We have developed the theoretical model for our system. In spite of the difference in the circuit design and qubit-resonator coupling mechanism, we obtain similar results in our system as in previously studied multimode circuit QED systems \cite{GelyPRB2017, MalekakhlaghPRL2017,Parra-RodriguezIOPSci2018}.
In particular, a cut-off effect arises naturally, such that the coupling strength between the qubit and the different modes is suppressed for high-frequency modes.
This cut-off effect prevents the qubit frequency from being suppressed to zero, which would happen if the qubit were coupled to an infinite number of resonator modes with no cut-off in the coupling strength.
We also derived a formula for the total amount of the Lamb shift induced by all the modes in a multimode resonator.

\section*{Results}

\begin{figure}[ht]
\includegraphics[scale=1.0]{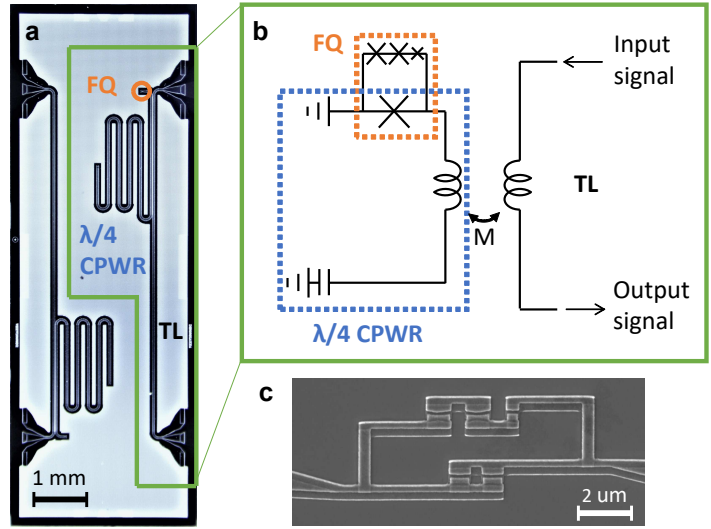}
\caption{
(\textbf{a}) Microscope image of the chip.
The chip contains two sets of independent circuit QED systems.
Each of the systems is inductively coupled to an independent TL.
(\textbf{b}) Model of a single circuit QED system.
The CPWR is inductively coupled to the TL through the mutual inductance M.
The spectroscopy signal is input from one side of the TL and the output signal transmitted to the other side is measured.
(\textbf{c}) SEM image of the qubit loop.
There are three Josephson junctions in the upper part of the loop as the conventional ones for a FQ and two Josephson junctions in the lower part of the loop acting as the coupling junction.
}
\label{fig:experimentcircuitdiagram}
\end{figure}

Figure \ref{fig:experimentcircuitdiagram} shows the schematics of our circuit QED system.
The CPWR is coupled to a transmission line (TL) through the mutual inductance (M in Fig. \ref{fig:experimentcircuitdiagram}\textbf{b}). The spectroscopy microwave signal is applied through the TL, and the transmitted signal is measured.
The FQ is placed at the short end of the $\lambda/4$ CPWR to couple with all the modes and a Josephson junction, which works as a simple inductance $L_{\rm{c}}$, is placed at the shared edge to enhance the coupling strength to achieve DSC \cite{NiemczykNPhys2010}.
Note that in this particular design, the coupling Josephson junction is split into parallel ones due to the difficulty to fabricate a large one, which results in a similar effect of a SQUID.
Due to the area difference between the FQ loop and the SQUID loop, the frequency modulation by the coupling SQUID is distinguishable from the modulation by the qubit.

\begin{figure}[ht]
\includegraphics[scale=0.9]{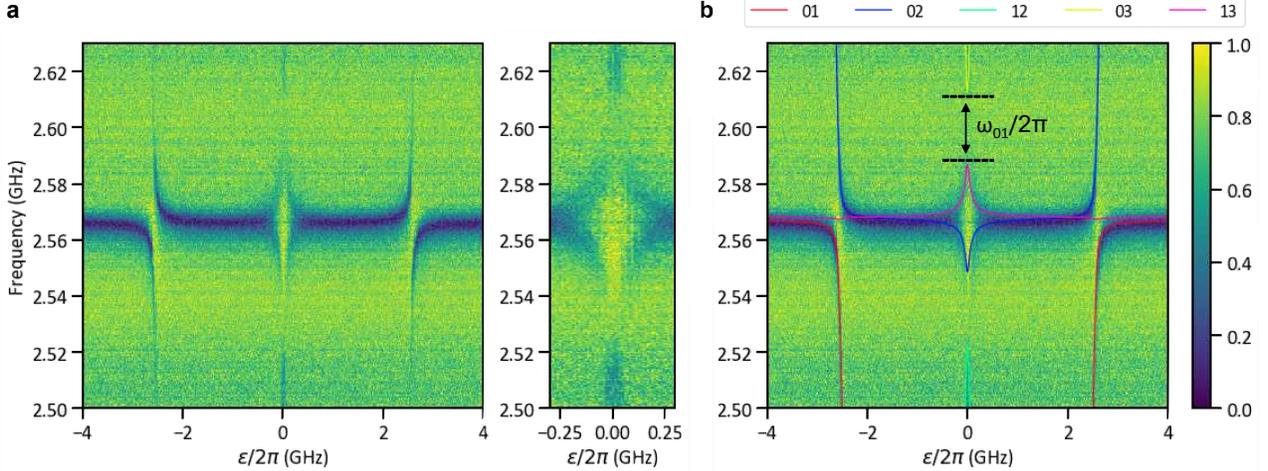}
\caption{
(\textbf{a}) Measured transmission spectrum around the CPWR fundamental mode frequency.
The wide panel shows the spectrum over a relatively wide range of flux bias values, while the narrow panel shows a close-up around the symmetry point. 
The x-axis is the flux bias, relative to the symmetry point, in frequency units. The y-axis is the probe frequency.
(\textbf{b}) Same spectrum along with theory curves calculated using Eq. (\ref{eq:totalHamiSM}) and the optimized fitting parameters: $\omega_1/2\pi = 2.57$ GHz, $\Delta_0^{'}/2\pi = 0.147$ GHz and $g_1/2\pi = 2.39$ GHz.
The labels of the theory curves indicate the transitions between pairs of energy eigenstates.
For example, the pair 01 means the transition from the ground state $\ket{0}$ to the first excited state $\ket{1}$ of the system. 
The frequency $\omega_{01}$ is the renormalized qubit frequency calculated from the theory curve.
The color bar shows the normalized amplitude of the measured output signal.
}
\label{fig:spectrafit}
\end{figure}

The transmission spectrum around the fundamental mode frequency is shown in Fig. \ref{fig:spectrafit}.
For the fitting of the spectrum, we used the single-mode quantum Rabi Hamiltonian
\begin{equation}
    \mathcal{H}_1 = -\frac{1}{2} (\Delta_0^{\prime} \sigma_{\rm{x}} + \varepsilon \sigma_{\rm{z}})
    + \omega_1 a_1^{\dagger}a_1 + g_1 \sigma_{\rm{z}} (a_1 + a_1^{\dagger}).
    \label{eq:totalHamiSM}
\end{equation}
Note that we take $\hbar = 1$.
Here, $\Delta_0^{\prime}$ is a partially renormalized qubit energy. Specifically, since there are a large number of modes in the CPWR, each mode contributes to the renormalization, i.e.~the Lamb shift, of the qubit energy. As a result, the qubit energy $\Delta_0^{\prime}$ that should be used in the single-mode QRM Hamiltonian will not be the bare qubit energy $\Delta_0$, but rather a smaller value that is renormalized by the qubit's interaction with all the modes except the fundamental mode.
The parameter $\varepsilon$ is the flux bias between the two qubit persistent current states, $\ket{R}$ and $\ket{L}$, defined such that the symmetry point is at $\varepsilon=0$.
$\omega_1$ is the mode energy of the fundamental mode in the CPWR, and $g_1$ is the coupling energy between the qubit and the fundamental mode.
$\sigma_{\rm{x}}$ and $\sigma_{\rm{z}}$ are the qubit's Pauli operators, and $a^{\dagger}$ and $a$ are the fundamental mode's creation and annihilation operators, respectively.

By fitting the spectrum produced by the Hamiltonian in Eq. (\ref{eq:totalHamiSM}) to the measured spectrum, we obtained the circuit parameters $\omega_1/2\pi = 2.57$ GHz, $\Delta_0^{'}/2\pi = 0.147$ GHz and $g_1/2\pi = 2.39$ GHz.
The fully renormalized qubit energy, i.e.~renormalized by all the CPWR modes, $\Delta$ is given by the experimentally observable transition frequency $\omega_{01}$ at the optimal bias point ($\varepsilon = 0$).
Here $\omega_{ij}$ is the frequency of the transition $\ket{i} \leftrightarrow \ket{j}$.
Two relatively minor complications arise in the measurement of $\omega_{01}$. First, $\omega_{01}$ at $\epsilon = 0$ is outside the measurable frequency range of our experimental setup, which is roughly 2-8 GHz. Although $\omega_{01}$ cannot be measured directly, it can be calculated straightforwardly by taking the difference $\omega_{03}-\omega_{13}$, or alternatively $\omega_{02}-\omega_{12}$. The second complication is that, because of state symmetry and selection rules, the 03 and 12 transitions are most clearly visible at $\varepsilon=0$, while the 13 and 02 transitions are forbidden at $\varepsilon=0$.\cite{YoshiharaNPhys2017} As a result, we do not observe spectral lines corresponding to the 13 and 02 transitions exactly at $\varepsilon=0$. Despite this complication, the values of $\omega_{13}$ and $\omega_{02}$ at $\varepsilon=0$ can be obtained straightforwardly from the fitting curves that match the observed spectrum very well for $\varepsilon \neq 0$. This fitting yields $\Delta/2\pi = \omega_{01}/2\pi = 26$ MHz. Note also that it is possible in principle to measure $\omega_{01}$ using two-tone spectroscopy, in which case no difficulties arise in relation to selection rules \cite{YoshiharaPRL2018}.

\begin{figure}[ht]
\includegraphics[scale=1.2]{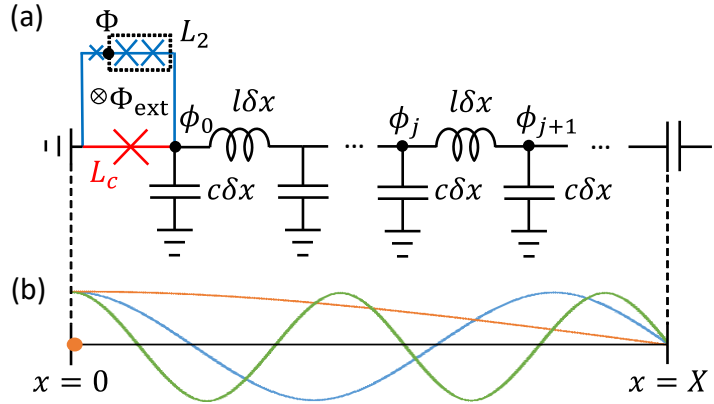}
\caption{\label{fig:epsart}
(\textbf{a}) Theoretical model for the circuit QED system.
The loop formed by the blue and red segments is the FQ loop, while the black and red parts form the CPWR.
The red part is the edge that is shared by the FQ and the CPWR with a Josephson junction inserted to achieve the DSC regime.
(\textbf{b}) Lowest three modes in the CPWR.
The color of the lines indicates the fundamental (orange), the second (blue), and the third (green) modes.
The orange dot indicates the location of the qubit where the qubit can couple to all modes in the CPWR.
Note that the scales of the two figures are different.
}
\label{fig:theorycircuitdiagram}
\end{figure}

To calculate the Lamb shift induced by all the modes in the resonator, we used the theoretical framework that describes the natural cut-off of the coupling for high-frequency modes.
We start by considering the circuit QED system shown in Fig. \ref{fig:theorycircuitdiagram}\textbf{a} which serves as a theoretical model for the experimental setup (Fig. \ref{fig:experimentcircuitdiagram}) without the TL.
Figure \ref{fig:theorycircuitdiagram}\textbf{b} illustrates the electric current amplitude profiles of the lowest three modes (the fundamental, second, and third modes) in the CPWR.
The qubit at the short end ($x=0$) couples to all the modes in the resonator, because the amplitudes of all the modes take their maximum values at the short end, keeping in mind that the coupling to the qubit alters this picture and suppresses the mode amplitudes at the point of contact.

Approximating the coupling Josephson junction as a linear inductance $L_{\rm{c}}$ and the two big FQ junctions together as a linear inductance $L_{\rm{2}}$ (respectively the red one and two blue ones in Fig. \ref{fig:theorycircuitdiagram}\textbf{a} for simplicity, the Lagrangian of this circuit can be expressed as
\begin{eqnarray}
    \mathcal{L} = \frac{1}{2} C_{\rm{q}} \dot{\Phi}^2 -  U_{\rm{q}}(\Phi,\Phi_{\rm{ext}})
    + \sum_{j=0}^{N} \left( \frac{1}{2} c\delta x \dot{\phi}_j^2 - \frac{1}{2l\delta x} \left( \phi_j - \phi_{j+1} \right)^2 \right)
    - \frac{1}{2L_{\rm{c}}} \phi_0^2 - \frac{1}{2L_2} \left( \Phi - \phi_0 \right)^2.
    \label{eq:circuitLagrangian}
\end{eqnarray}
Here, $\Phi$ is the flux (or phase) variable at the node at the right side of the small junction in the qubit loop and $\phi_j$ is the phase at the node $j$ ($j=0,1,2,...,N$) of the CPWR.,
$C_{\rm{q}}$ is the capacitance of the small blue junction in Fig. \ref{fig:theorycircuitdiagram}\textbf{a} and $U_{\rm{q}}(\Phi,\Phi_{\rm{ext}})=-\alpha E_{\rm{J}} \cos \left\{ {2 \pi (\Phi- \Phi_{\rm{ext}})/\Phi_0 } \right\}$ is the effective Josephson potential for the small Josephson junction.
$\alpha$ is the area ratio of the small junction, $E_{\rm{J}}=I_{\rm{c}} \Phi_0 / (2\pi)$ is the Josephson energy of the large junctions, $I_{\rm{c}}$ is the critical current of the Josephson junction, and $\Phi_0=h/(2e)$ is the superconducting flux quantum.
$c$ and $l$ are, respectively, the capacitance and the inductance per unit length of the resonator, which yields the capacitance $c\delta x $ and the inductance $l\delta x$ for each section $\delta x$ in the resonator.
Using the Legendre transformation, $Q = \partial \mathcal{L} / \partial \dot{\Phi} = C_{\rm{q}} \dot{\Phi}$ and $q_j = \partial \mathcal{L} / \partial \dot{\phi_j} = c \dot{\phi_j}$, the Hamiltonian of the circuit can be obtained as
\begin{eqnarray}
    \mathcal{H} = \frac{1}{2C_{\rm{q}}}Q^2 + U_{\rm{q}}(\Phi,\Phi_{\rm{ext}})
    + \sum_{j=0}^N \left( \frac{1}{2c\delta x}q_j^2 + \frac{1}{2l\delta x}(\phi_j - \phi_{j+1})^2 \right)
    + \frac{1}{2L_{\rm{c}}} \phi_0^2 + \frac{1}{2L_{2}}(\Phi - \phi_0)^2.
    \label{eq:circuitHami}
\end{eqnarray}

By ignoring the qubit term, represented by the variables $\Phi$ and $Q$, the equations of motion yield
\begin{equation}
    \omega_{\rm{cutoff}} = \frac{Z_0}{L_{\rm{c}2}}.
    \label{eq:cutofffrequencyMain}
\end{equation}
with consideration of the boundary conditions at $x = 0$ and $x = X$ (more details in the Methods section).
Here, $Z_0 = \sqrt{l/c}$ is the characteristic impedance of the CPWR and $L_{\rm{c}2} = L_{\rm{c}} L_2 / (L_{\rm{c}} + L_2)$.
Since $L_{\rm{c}} \ll L_2$, the inductance $L_{\rm{c}2}$ can be approximated as $L_{\rm{c}}$.
This result is consistent with previous results in the literature \cite{MalekakhlaghPRL2017, GelyPRB2017,Parra-RodriguezIOPSci2018}, although a different circuit design and capacitive coupling between a transmon qubit and a CPWR were considered in those studies.

\begin{figure}[ht]
\includegraphics[scale=0.9]{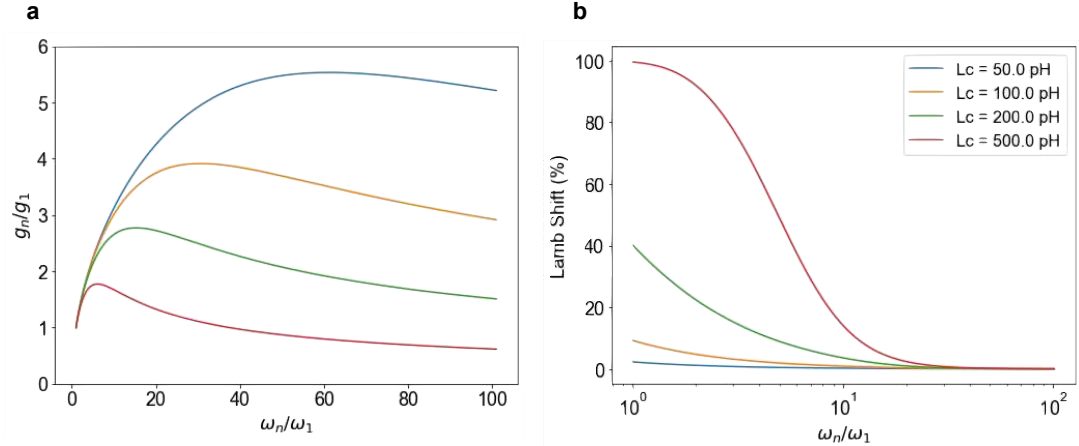}
\caption{\label{fig:lambshift-gn}
(\textbf{a}) Normalized coupling strength $g_n/g_1$ as a function of mode frequency.
The maximum in $g_n$ occurs at $\omega_n=\omega_{\rm{cutoff}}$.
The plots show that larger $L_c$ yields smaller $\omega_{\rm{cutoff}}$ as described in Eq. (\ref{eq:cutofffrequencyMain}).
The colors of the lines follow the legend in (\textbf{b}).
(\textbf{b}) Lamb shift induced by the individual modes.
The Lamb shift induced by one mode is calculated by assuming coupling between the qubit and only that one mode. The shift is therefore calculated relative to $\Delta_0$. In other words, the relative Lamb shift is given by $(\Delta_0 - \Delta)/\Delta_0$.
The decrease in this Lamb shift with increasing mode frequency occurs both with and without the cutoff effect.
The cutoff effect keeps the total Lamb shift below 100\%, while it approaches 100\% in the absence of a high-frequency cutoff.
This effect cannot be straightforwardly inferred from the line shapes in the figure.
Note that the plot in (\textbf{b}) uses a logarithmic scale for the x-axis.
To generate the plots in this figure, we use the design circuit parameters: $L=1.93$ nH, $L_2=823$ pH, $\alpha=0.46$, and $E_{\rm{J}}=397$ GHz.}
\end{figure}

From the mode frequencies and current profile functions, we calculated the zero-point fluctuation of the current $I_n^{\rm{zpf}}$ for the $n$th mode to obtain the coupling strength between the qubit and the $n$th mode as $g_n = L_{\rm{c}} I_{\rm{q}} I_n^{\rm{zpf}}$, where $I_{\rm{q}}$ is the qubit persistent current \cite{YoshiharaNPhys2017, YoshiharaSciRep2022}.
As explained in the Methods section, when $\omega_{\rm{cutoff}}\gg\omega_1$, this formula for $g_n$ yields
\begin{equation}
    \label{eq:gnMain}
    g_n = g_1 \times \sqrt{\frac{ \frac{\omega_n}{\omega_{1}}}{1 + \left(\frac{\omega_n}{\omega_{\rm{cutoff}}} \right)^2}}.
\end{equation}
Here $\omega_n$ is the bare frequency of the $n$th mode.
For the low-frequency modes ($\omega_n \ll \omega_{\rm{cutoff}}$), $g_n \propto \sqrt{\omega_n}$, and for the high-frequency modes ($\omega_n \gg \omega_{\rm{cutoff}}$), $g_n \propto 1 / \sqrt{\omega_n}$.
Figure \ref{fig:lambshift-gn}\textbf{a} shows the behavior of $g_n$ as a function of mode frequency for a few different values of the coupling inductance $L_{\rm{c}}$. 
For this figure, we used the design circuit parameters: the total inductance of the resonator $L=1.93$ nH, $L_2=823$ pH, $\alpha=0.46$, and $E_{\rm{J}}=397$ GHz.

Now we derive the formula for the Lamb shift caused by the multiple modes in the CPWR using the formula for $g_n$.
For the single-mode case, as described by Eq. (\ref{eq:totalHamiSM}), the renormalized qubit energy $\Delta$ can be expressed as a function of the bare qubit energy $\Delta_0$, the resonant frequency $\omega$ of the resonator mode, and the coupling strength $g$:
\begin{equation}
    \Delta = \Delta_0 \exp (-2g^2/\omega^2)
    \label{eq:singleLS}
\end{equation}
which is valid when $\Delta_0 \ll \omega$.\cite{AshhabPRA2010}
From the bare and renormalized values, the relative Lamb shift is calculated as $(\Delta_0 - \Delta)/\Delta_0$.
When coupled to a multi-mode resonator, the qubit energy will be renormalized by all the modes, and the fully renormalized qubit energy can be expressed using a product of exponential terms corresponding to all the modes:
\begin{eqnarray}
    \Delta = \Delta_0 \prod_{n}{\exp (-2g_n^2/\omega_n^2)} = \Delta_0 \exp \left(-2\sum_n g_n^2/\omega_n^2 \right).
    \label{eq:multiLS}
\end{eqnarray}
It is worth reiterating here that, since we keep only the fundamental mode in Eq. (\ref{eq:totalHamiSM}), we encounter a number of qubit energy values that correspond to different levels of renormalization, as described by the relations
\begin{equation}
    \Delta_0^{\prime} = \Delta_0 \prod_{n \neq 1}{\exp (-2g_n^2/\omega_n^2)},
\end{equation}
and
\begin{equation}    
    \Delta = \Delta_0^{\prime} {\exp (-2g_1^2/\omega_1^2)}.
\end{equation}
The bare qubit energy is $\Delta_0$, while $\Delta_0^{\prime}$ includes renormalization by the coupling to the high-frequency modes, and $\Delta$ is the fully renormalized value.

Using Eq. (\ref{eq:gnMain}) for $g_n$ and taking into consideration that the mode frequencies in a $\lambda/4$ CPWR are odd multiples of the fundamental mode frequency, Eq. (\ref{eq:multiLS}) can be transformed into
\begin{equation}
    \Delta = \Delta_0 \exp \left\{-2 \left(\frac{g_1}{\omega_1} \right)^2 \sum_{n=1,3,...} \frac{1}{n\left( 1 + \frac{n^2}{n_{\rm{cutoff}}^2} \right) } \right\}.
    \label{eq:multiLScutoff}
\end{equation}
Here $n_{\rm{cutoff}} = \omega_{\rm{cutoff}}/\omega_1$. The sum inside the exponential in Eq. (\ref{eq:multiLScutoff}) is larger than one, which increases the total Lamb shift compared to the single-mode value.
Figure \ref{fig:lambshift-gn}\textbf{b} shows the Lamb shift that each mode induces in $\Delta_0$.
The Lamb shift values plotted in Fig. \ref{fig:lambshift-gn}\textbf{b} are calculated using the formula $\left(\Delta_0 - \Delta_n\right)/\Delta_0$, where $\Delta_n$ is the renormalized qubit energy if the qubit coupled to the $n$th mode only.
Using the software package Mathematica, we find that the sum in Eq. (\ref{eq:multiLScutoff}) approaches $0.25(2\gamma+\rm{log}4)+0.5\rm{log} n_{\rm{cutoff}}$, where $\gamma$ is Euler's constant, in the limit $n_{\rm{cutoff}}\to\infty$. Put differently, the sum can be very well approximated by the simple formula $0.635 + 0.5 \rm{log} n_{\rm{cutoff}}$, provided that $n_{\rm{cutoff}} \gg 1$. The replacement of the sum by this simple formula illustrates the important point that we do not need to characterize the coupling between the qubit and every resonator mode individually. The system parameters that we obtained using the single-mode model including only the fundamental mode allows us to estimate the total Lamb shift contributed by all the modes.
Using the value of the impedance $Z_0 = 50 \ \Omega$ of the resonator and the coupling inductance $L_{\rm{c}} = 231$ pH, we estimate that $\omega_{\rm{cutoff}}/2\pi \sim 34.4$ GHz, or alternatively $n_{\rm{cutoff}}=13.2$.
Note that $L_{\rm{c}}$ is calculated from the difference between the designed bare fundamental mode frequency and the measured fundamental mode frequency at the zero flux point.
In this case, the sum in Eq. (\ref{eq:multiLScutoff}) yields 1.93.
Using the system parameters obtained from the fitting, we find that the total Lamb shift in our multimode system is 96.5\%.
Using this result and the measured $\Delta/2\pi = 26$ MHz, the bare qubit energy can be calculated as $\Delta_0/2\pi \sim 732$ MHz.

\section*{Discussion}

We obtained experimental and theoretical results pertaining to the extremely large Lamb shift that occurs in a DSC multimode circuit QED system containing a superconducting FQ and a CPWR.
We used the single-mode quantum Rabi Hamiltonian to fit the measured spectrum near the fundamental mode frequency to obtain the circuit parameters.
The measured spectrum shows that our system is in the DSC regime.
The single-mode QRM model was used to obtain the system parameters. 
The fitting gave the fully renormalized qubit energy as $\Delta/2\pi = 26$ MHz.
Since we cannot turn off the coupling between the qubit and the resonator, we cannot isolate the qubit from the resonator and directly measure the bare qubit energy.
We can only estimate the bare qubit energy based on the experimentally measured system parameters. However, considering how well the theoretical spectrum fits the experimentally observed spectrum, we can have a good amount of confidence in the validity of the model and hence in our estimate for the bare qubit energy.
To deal with the high-frequency modes, we considered the natural cut-off in the coupling strength $g_n$ that occurs at high frequencies, and we calculated $g_n$ as a function of $\omega_{\rm{cutoff}}$. The dependence of $g_n$ on $\omega_{\rm{cutoff}}$ in our system is similar to that obtained with different circuit designs.
Using the formula for $g_n$, we derived the formula for the renormalized qubit energy $\Delta$, which yields the Lamb shift.
Our theoretical results demonstrate how $\Delta$ survives the renormalization by the Lamb shift and remains finite even in the case of strong coupling to an infinite number of modes in the CPWR.
The total Lamb shift in our system was calculated to be 96.5\%, which is larger than most values measured previously using single-mode resonators.

\section*{Methods}
\label{sec:method}

\subsection* {\label{app:delivationcutoff}Derivation of the cutoff frequency}
To find the modes in the CPWR, a standard approach is to ignore qubit terms (i.e. $\Phi$) in the first-order and second-order equations of motion, which gives
\begin{eqnarray}
    \frac{\partial^2 \phi}{\partial t^2} &=& \frac{1}{cl} \frac{\partial^2 \phi}{\partial x^2} \\
    \left. \left( \frac{1}{X^2cl} \frac{\partial \phi}{\partial x} - \frac{\phi}{XcL_{\rm{c}}} - \frac{\phi}{XcL_2} \right) \right|_{x=0} &=& 0 \\
    \left. \frac{\partial \phi}{\partial x}\right|_{x=X} &=& 0.
\end{eqnarray}
$x$ indicates the location in the CPWR with the two boundaries, $x=0$ and $x=X$.
By substituting the solution $\phi(x,t)=e^{i \omega t}u(x)$ with a temporal frequency $\omega$ and time $t$ in these equations, we obtain
\begin{eqnarray}
    \label{eq:secmotioneq} - \omega^2 u(x) &=& \frac{1}{cl} \frac{\partial^2 u}{\partial x^2} \\
    \label{eq:bound0}      \left. \left( \frac{1}{cl} \frac{\partial u}{\partial x} - \frac{u}{cL_{\rm{c}2}} \right) \right|_{x=0} &=& 0 \\
    \label{eq:boundX}      \left. \frac{\partial u}{\partial x} \right|_{x=X} &=& 0
\end{eqnarray}
with $L_{\rm{c}2} = L_{\rm{c}} L_2 / (L_{\rm{c}} + L_2)$.
The solution of Eq. (\ref{eq:secmotioneq}) can be written as $u(x) = u_{\rm{c}} \cos (kx) + u_{\rm{s}} \sin (kx)$, where $k = \sqrt{\omega^2 cl}$. The boundary conditions Eq. (\ref{eq:bound0}) and Eq. (\ref{eq:boundX}) now give us $k \tan (kX) = l / L_{\rm{c}2}$, or equivalently,
\begin{equation}
    \omega \tan (\omega \sqrt{cl} X) = \omega_{\rm{cutoff}}
    \label{eq:frequencycondition}
\end{equation}
where
\begin{equation}
    \omega_{\rm{cutoff}} = \frac{Z_0}{L_{\rm{c}2}}.
    \label{eq:cutofffrequency}
\end{equation}
Here, $Z_0 = \sqrt{l/c}$ is the characteristic impedance of the CPWR.

\subsection* {\label{app:zpf}Derivation of the zero-point fluctuation of the current}
For the low-frequency modes whose frequency is less than $\omega_{\rm{cutoff}}$, the solutions of Eq. (\ref{eq:frequencycondition}) must have $\tan (kX) \gg 1$, which allows us to make the approximation that $kX$ is close to $n \pi - \pi/2$, where n is a positive integer starting at $n=1$ for the fundamental mode.
After defining $k \tilde{X} = n \pi - \pi/2 - kX$, the solutions can be written as
\begin{eqnarray}
    k \tilde{X} = n \pi - \frac{\pi}{2} \nonumber - X\sqrt{cl} \omega_{\rm{cutoff}} \cot \left( n \pi - \frac{\pi}{2} - k \tilde{X} \right).
\end{eqnarray}
Making the approximation that $\cot \left( n \pi - \pi / 2 - \delta \right) \approx \delta$ gives the first-order approximation in $\omega/ \omega_{\rm{cutoff}}$ as
\begin{equation}
    k \tilde{X} \approx \frac{n \pi - \frac{\pi}{2}}{1+X\sqrt{cl} \omega_{\rm{cutoff}}}
\end{equation}
and therefore
\begin{eqnarray}
    kX \approx n \pi - \frac{\pi}{2} - \frac{\left( n \pi - \frac{\pi}{2} \right)}{X\sqrt{cl} \omega_{\rm{cutoff}}}
    = \left( n \pi - \frac{\pi}{2} \right) \times \left( 1 - \frac{L_{\rm{c}2}}{Xl} \right)
\end{eqnarray}
which gives us the frequency of the fundamental mode $\omega_1 = \pi / 2 X\sqrt{cl}$, as expected for a quarter-wavelength CPWR.

The energy $E_n$ of each mode is proportional to $u_{\rm{c}}^2$:
\begin{equation}
    E_n = \frac{Xk^2u_{\rm{c}^2}}{2l}.
    \label{eq:ModeEnergy}
\end{equation}
In the ground state of each mode, the energy should be $E_n = \omega_n / 2$, which therefore gives the zero-point fluctuations in the mode variable
\begin{equation}
    u_{\rm{c, rms}} = \sqrt{ \frac{1}{Xc \omega_n}}.
    \label{eq:ModeFluctuation}
\end{equation}
Now we can calculate the zero-point fluctuations in the current as
\begin{eqnarray}
    I_n^{\rm{zpf}} = \frac{1}{Xl} \left| \frac{\partial u}{\partial x} \right|_{x=0,rms}
    = \sqrt{ \frac{1}{Xc \omega_n}} \frac{k \sin (kX)}{Xl}
    = \sqrt{\frac{1}{Xl}} \sqrt{ \frac{\omega_n}{1 + \left( \frac{\omega_n}{\omega_{\rm{cutoff}}} \right)^2}}.
    \label{eq:Izpf}
\end{eqnarray}
Considering the formula $g_n = L_{\rm{c}} I_{\rm{q}} I_n^{\rm{zpf}}$, treating $L_{\rm{c}}$ and $I_{\rm{q}}$ as constants and making the approximation $\omega_{\rm{cutoff}}\gg\omega_1$, we obtain Eq. (\ref{eq:gnMain}).


\begin{thebibliography}{99}

\bibitem{NielsenChuang10th} Nielsen, M. A. and Chuang, I. L. Quantum Computation and Quantum Information: 10th Anniversary Edition (Cambridge University Press, 2010).

\bibitem{KimbleNature2008} Kimble, H. J. The quantum internet. Nature 453, 1023 (2008).

\bibitem{DevoretSchoelkopfScience2013} Devoret, M. H. and Schoelkopf, R. J. Superconducting circuits for quantum information: An outlook. Science 339, 1169 (2013).

\bibitem{LambPhysRev1947} Lamb, W. E. and Retherford, R. C. Fine structure of the hydrogen atom by a microwave method. Phys. Rev. 72, 241–243 (1947).

\bibitem{BrunePRL1994} Brune, M. et al. From lamb shift to light shifts: Vacuum and subphoton cavity fields measured by atomic phase sensitive detection. Phys. Rev. Lett. 72, 3339–3342 (1994).

\bibitem{FragnerScience2008} Fragner, A. et al. Resolving vacuum fluctuations in an electrical circuit by measuring the lamb shift. Science 322, 1357–1360 (2008).

\bibitem{Forn-DiazRevModPhys2019} Forn-Díaz, P., Lamata, L., Rico, E., Kono, J. and Solano, E. Ultrastrong coupling regimes of light-matter interaction. Rev. Mod. Phys. 91, 025005 (2019).

\bibitem{KockumNRevPhys2019} Frisk Kockum, A., Miranowicz, A., De Liberato, S., Savasta, S. and Nori, F. Ultrastrong coupling between light and matter. Nat. Rev. Phys. 1, 025005 (2019).

\bibitem{YoshiharaNPhys2017} Yoshihara, F. et al. Superconducting qubit–oscillator circuit beyond the ultrastrong-coupling regime. Nat. Phys. 13, 44 (2017).

\bibitem{YoshiharaPRA2017} Yoshihara, F. et al. Characteristic spectra of circuit quantum electrodynamics systems from the ultrastrong- to the deep-strong-coupling regime. Phys. Rev. A 95, 053824 (2017).

\bibitem{AnapparaPRB2009} Anappara, A. A. et al. Signatures of the ultrastrong light-matter coupling regime. Phys. Rev. B 79, 201303 (2009).

\bibitem{GambinoACSPhoto2014} Gambino, S. et al. Exploring light–matter interaction phenomena under ultrastrong coupling regime. ACS Photonics 1, 1042 (2014).

\bibitem{BenzScience2016} Benz, F. et al. Single-molecule optomechanics in ``picocavities''. Science 354, 726–729 (2016).

\bibitem{BayerACSPhoto2017} Bayer, A. et al. Terahertz light–matter interaction beyond unity coupling strength. Nano Lett. 17, 6340 (2017).

\bibitem{VrajitoareaarXiv2022} Vrajitoarea, A. et al. Ultrastrong light-matter interaction in a photonic crystal. arXiv:2209.14972 (2022).

\bibitem{BlaisPRA2004} Blais, A., Huang, R.-S., Wallraff, A., Girvin, S. M. and Schoelkopf, R. J. Cavity quantum electrodynamics for superconducting electrical circuits: An architecture for quantum computation. Phys. Rev. A 69, 062320 (2004).

\bibitem{WallraffNature2004} Wallraff, A. et al. Strong coupling of a single photon to a superconducting qubit using circuit quantum electrodynamics. Nature 431, 162–167 (2004).

\bibitem{ChiorescuNature2004} Chiorescu, I. et al. Coherent dynamics of a flux qubit coupled to a harmonic oscillator. Nature 431, 159 (2004).

\bibitem{BourassaPRA2009} Bourassa, J. et al. Ultrastrong coupling regime of cavity qed with phase-biased flux qubits. Phys. Rev. A 80, 032109 (2009).

\bibitem{NiemczykNPhys2010} Niemczyk, T. et al. Circuit quantum electrodynamics in the ultrastrong-coupling regime. Nat. Phys. 6, 772–776 (2010).

\bibitem{LizuainPRA2010} Lizuain, I., Casanova, J., García-Ripoll, J. J., Muga, J. G. and Solano, E. Zeno physics in ultrastrong-coupling circuit qed.
Phys. Rev. A 81, 062131 (2010).

\bibitem{Forn-DiazPRL2010} Forn-Díaz, P. et al. Observation of the bloch-siegert shift in a qubit-oscillator system in the ultrastrong coupling regime. Phys. Rev. Lett. 105, 237001 (2010).

\bibitem{FelicettiPRA2014} Felicetti, S., Romero, G., Rossini, D., Fazio, R. and Solano, E. Photon transfer in ultrastrongly coupled three-cavity arrays. Phys. Rev. A 89, 013853 (2014).

\bibitem{Forn-DiazNatPhys2017} Forn-Díaz, P. et al. Ultrastrong coupling of a single artificial atom to an electromagnetic continuum in the nonperturbative regime. Nat. Phys. 13, 39 (2017).

\bibitem{SundaresanPRX2015} Sundaresan, N. M. et al. Beyond strong coupling in a multimode cavity. Phys. Rev. X 5, 021035 (2015).

\bibitem{LiuNatPhys2017} Liu, Y. and Houck, A. A. Quantum electrodynamics near a photonic bandgap. Nat. Phys. 13, 48 (2017).

\bibitem{BosmannpjQI2017} Bosman, S. J. et al. Multi-mode ultra-strong coupling in circuit quantum electrodynamics. npj Quantum Inf. 3, 46 (2017).

\bibitem{MartineznpjQI2019} Martínez, J. P. et al. A tunable josephson platform to explore many-body quantum optics in circuit-qed. npj Quantum Inf. 5, 19 (2019).

\bibitem{KoshinoPRResearch2022} Koshino, K., Shitara, T., Ao, Z. and Semba, K. Deterministic three-photon down-conversion by a passive ultrastrong cavity-qed system. Phys. Rev. Res. 4, 013013 (2022).

\bibitem{YoshiharaPRL2018} Yoshihara, F. et al. Inversion of qubit energy levels in qubit-oscillator circuits in the deep-strong-coupling regime. Phys. Rev. Lett. 120, 183601 (2018).

\bibitem{YoshiharaSciRep2022} Yoshihara, F., Ashhab, S., Fuse, T., Bamba, M. and Semba, K. Hamiltonian of a flux qubit-lc oscillator circuit in the deep–strong-coupling regime. Sci. Rep. 12, 6764 (2022).

\bibitem{GelyPRB2017} Gely, M. F. et al. Convergence of the multimode quantum rabi model of circuit quantum electrodynamics. Phys. Rev. B 95, 245115 (2017).

\bibitem{MalekakhlaghPRL2017} Malekakhlagh, M., Petrescu, A. and Türeci, H. E. Cutoff-free circuit quantum electrodynamics. Phys. Rev. Lett. 119, 073601 (2017).

\bibitem{Parra-RodriguezIOPSci2018} Parra-Rodriguez, A., Rico, E., Solano, E. and  Egusquiza, I. L. Quantum networks in divergence-free circuit QED. Quant. Sci. Technol. 3, 024012 (2018).

\bibitem{AshhabPRA2010} Ashhab, S. and Nori, F. Qubit-oscillator systems in the ultrastrong-coupling regime and their potential for preparing nonclassical states. Phys. Rev. A 81, 042311 (2010).

\end{thebibliography}

\section*{Acknowledgments}

We would like to thank Adrian Parra-Rodriguez for the useful discussions.
This work was supported by Japan Science and Technology Agency Core Research for Evolutionary Science and Technology (Grant No. JPMJCR1775).

\section*{Author contributions statement}

Z.A., F.Y., and K.S. conceived the experiment.
Z.A., F.Y., and T.F. designed the samples and performed the measurements.
K.K. and S.S. fabricated the samples.
S.A. performed the theoretical calculations.
Z.A., S.A., F.Y., and T.F. analyzed the measurement data.
Z.A. and S.A. wrote the manuscript with feedback from all authors.
T.A. and K.S. supervised the project.

\section*{Data Availability}

The datasets generated and/or analyzed during the current study are available from the corresponding authors on reasonable request.

\end{document}